\documentclass{ws-procs9x6} 
\newcommand{\beq}{\begin{eqnarray}}
\newcommand{\eeq}{\end{eqnarray}}
\newcommand{\beqnn}{\begin{eqnarray*}}
\newcommand{\eeqnn}{\end{eqnarray*}}

\newcommand{\Tr}{\mathop{\mathrm{Tr}}}

\newcommand{\tp}[1]{\:{}^{\mathrm{t}}#1}

\newcommand{\CC}{\mathbf{C}}

\newcommand{\ZZ}{\mathbf{Z}}
\newcommand{\NN}{\mathrm{N}}
\newcommand{\bst}{\boldsymbol{t}}
\newcommand{\bsx}{\boldsymbol{x}}
\newcommand{\bsy}{\boldsymbol{y}}
\newcommand{\bsu}{\boldsymbol{u}}
\newcommand{\bsv}{\boldsymbol{v}}

\newcommand{\tbar}{\bar{t}}
\newcommand{\zerobar}{\bar{0}}
\newcommand{\bstbar}{\bar{\bst}}

\newcommand{\calT}{\mathcal{T}}

\begin{document}

\title{KP and Toda tau functions in Bethe ansatz}
\author{Kanehisa Takasaki}
\address{Graduate School of Human and Environmental Studies\\
Kyoto University\\
Yoshida, Sakyo, Kyoto, 606-8501, Japan\\
takasaki@math.h.kyoto-u.ac.jp}

\begin{abstract}
Recent work of Foda and his group on a connection between 
classical integrable hierarchies (the KP and 2D Toda hierarchies) 
and some quantum integrable systems (the 6-vertex model with DWBC, 
the finite XXZ chain of spin $1/2$, the phase model on a finite chain, 
etc.) is reviewed.  Some additional information on this issue is 
also presented.   
\end{abstract}

\keywords{six-vertex model; XXZ model; 
domain wall boundary condition; Bethe ansatz; 
KP hierarchy; Toda hierarchy; tau function}

\bodymatter

\section{Introduction}

Searching for a connection between classical and quantum 
integrable systems is an old and new subject, 
occasionally leading to a breakthrough towards 
a new area of research. One of the landmarks in this sense 
is the quantum inverse scattering method, 
also known as the algebraic Bethe ansatz.  
Stemming from the classical inverse scattering method, 
the algebraic Bethe ansatz covers a wide class of 
integrable systems including solvable models 
of statistical mechanics on the basis of 
the Yang-Baxter equations \cite{YBE-book}. 
Moreover, remarkably, it was recognized later 
that a kind of classical integrable systems 
(discrete Hirota equations) show up in 
the so called nested Bethe ansatz \cite{Zabrodin96}.  

Recently a new connection was found by Foda and his group 
\cite{FWZ09a,FWZ09b,FS10,Zuparic09,Zuparic-thesis}. 
They observed that special solutions of 
the classical integrable hierarchies 
(the KP and 2D Toda hierarchies) are hidden 
in quantum (or statistical) integrable systems 
such as the 6-vertex model under 
the domain wall boundary condition (DWBC)\cite{FWZ09a}, 
the finite XXZ chain of spin $1/2$ \cite{FWZ09b,FS10},
and some other quantum integrable systems 
\cite{Zuparic09,Zuparic-thesis}.  
Their results are based on a determinant formula 
of physical quantities, namely, 
the Izergin-Korepin formula for the partition function 
of the 6-vertex model \cite{Korepin82,Izergin87,ICK91} 
and the Slavnov formula for the scalar product of Bethe states 
in the XXZ spin chain \cite{Slavnov89,KMT99}.  
Those formulae contain a set of free variables, 
and the determinant in the formula is divided 
by the Vandermonde determinant of these variables.  
Foda et al. interpreted the quotient of the determinant 
by the Vandermonde determinants as a tau function 
of the KP (or 2D Toda) hierarchy expressed 
in the so called ``Miwa variables''.  

In this paper, we review these results along with 
some additional information on this issue.  
We are particularly interested in the relevance 
of the 2-component KP (2-KP) and 2D Toda hierarchies.  
Unfortunately, this research is still in an early stage, 
and we cannot definitely say which direction 
this research leads us to.  A modest goal will be 
to understand the algebraic Bethe ansatz better 
in the perspective of classical integrable hierarchies. 

This paper is organized as follows.  
In Section 2, we start with a brief account of 
the notion of Schur functions that play a fundamental role 
in the theory of integrable hierarchies, and 
introduce the tau function of the KP, 2-KP and 
2D Toda hierarchies as a function of both 
the usual time variables and the Miwa variables.  
Section 3 deals with the partition function 
of the 6-vertex model with DWBC.  
Following the procedure of Foda et al., 
we rewrite the Izergin-Korepin formula 
into an almost rational form and show that 
a main part of the partition function 
can be interpreted as a KP tau function.  
Actually, the partition function allows 
two different interpretations that correspond 
to two choices of the Miwa variables.  
We examine a unified interpretation of 
the partition function as a tau function 
of the 2-KP (or 2D Toda) hierarchy.  
In Section 4, we turn to the finite XXZ chain 
of spin $1/2$, and present a similar interpretation 
to the scalar product of Bethe states (one of which 
depends on free variables) on the basis of the Slavnov formula.  
Section 5 is devoted to some other models 
including the phase model \cite{BIK97}, which is 
also studied by the group of Foda \cite{Zuparic09}.  
For those models, a determinant formula is known 
to hold for the scalar product of Bethe states 
both of which depend on free parameters \cite{KBI-book}. 
We consider a special case related to enumeration 
of boxed plane partitions.

\section{Tau functions}

\subsection{Schur functions}

Let us review the notion of Schur functions.  
We mostly follow the notations of Macdonald's book 
\cite{Macdonald-book}.

For $N$ variables $\bsx = (x_1,\ldots,x_N)$ 
and a partition $\lambda = (\lambda_1,\lambda_2,\ldots,\lambda_N)$ 
($\lambda_1 \ge \lambda_2 \ge\cdots\ge \lambda_N \ge 0$) 
of length $(\lambda) \le N$, the Schur function $s_\lambda(\bsx)$ 
can be defined by Weyl's character formula 
\beq
  s_\lambda(\bsx) 
  = \frac{\det(x_j^{\lambda_i-i+N})_{i,j=1}^N}{\Delta(\bsx)}, 
\label{Weyl}
\eeq
where $\Delta(\bsx)$ is the Vandermonde determinant 
\beqnn
  \Delta(\bsx) 
  = \det(x_j^{-i+N})_{i,j=1}^N 
  = \prod_{1\le i<j\le N}(x_i - x_j). 
\eeqnn

By one of the Jacobi-Trudi formulae, $s_\lambda(\bsx)$ 
can be expressed as a determinant of the form 
\beq
  s_\lambda(\bsx) = \det(h_{\lambda_i-i+j}(\bsx))_{i,j=1}^N, 
\label{Jacobi-Trudi}
\eeq
where $h_n(\bsx)$, $n = 0,1,2,\ldots$, 
are the completely symmetric functions 
\beqnn
  h_n(\bsx) 
  = \sum_{1\le k_1\le k_2 \le \cdots\le k_n \le N}
    x_{k_1}x_{k_2}\cdots x_{k_n} 
  \quad \mbox{for}\quad n \ge 1, \quad 
  h_0(\bsx) = 1. 
\eeqnn
The complete symmetric functions $h_n(\bsx)$ themselves 
can be identified with the Schur functions 
for partitions with a single part: 
\beqnn
  h_n(\bsx) = s_{(n)}(\bsx),\quad 
  (n) := (n,0,\ldots,0). 
\eeqnn
Another form of the Jacobi-Trudi formulae uses 
on the elementary symmetric functions $e_n(\bsx)$.  
Since we shall not use it in the following, 
its details are omitted here.  

The complete symmetric functions have 
the generating function
\beqnn
  \sum_{n=0}^\infty h_n(\bsx)z^n 
  = \prod_{k=1}^N(1 - x_kz)^{-1} 
  = \exp\left(- \sum_{k=1}^N\log(1 - x_kz)\right). 
\eeqnn
Since $\log(1 - x_kz)$ has a Taylor expansion of the form 
\beqnn
  \log(1 - x_kz) = - \sum_{n=1}^\infty \frac{x_k^n}{n}z^n, 
\eeqnn
this generating function can be rewritten as 
\beq
  \sum_{n=0}^\infty h_n(\bsx)z^n 
  = \exp\left(\sum_{n=1}^\infty t_nz\right), 
\label{hn(x)-genfun}
\eeq
where $t_n$'s are defined as 
\beq
  t_n = \frac{1}{n}\sum_{k=1}^N x_k^n. 
\label{t-x}
\eeq
In view of (\ref{Jacobi-Trudi}) and (\ref{hn(x)-genfun}), 
one can redefine the complete symmetric functions 
and the Schur functions as functions $h_n[\bst]$ 
and $s_\lambda[\bst]$ \footnote{These convenient notations 
are borrowed from Zinn-Justin's paper \cite{Zinn-Justin09}.} 
of $\bst = (t_1,t_2,\ldots)$, namely, 
\beq
  \sum_{n=0}^\infty h_n[\bst]z^n 
  = \exp\left(\sum_{n=1}^\infty t_nz^n\right)
\eeq
and
\beq
  s_\lambda[\bst] = \det(h_{\lambda_i-i+j}[\bst])_{i,j=1}^N, 
\label{Jacobi-Trudi[t]}
\eeq
where $\lambda$ is understood to be an infinite 
decreasing sequence $\lambda = (\lambda_1,\lambda_2,\ldots)$ 
with $\lambda_i = 0$ for all but a finite number of $i$'s, 
and $N$ is arbitrary integer greater than or equal 
to $l(\lambda) = \max\{i \mid \lambda_i \not= 0\}$. 
The right hand side of (\ref{Jacobi-Trudi[t]}) 
is independent of $N$, because the lower left block 
of the matrix therein for $i > l(\lambda)$ and 
$j \le l(\lambda)$ is zero and the lower right block 
for $i,j > l(\lambda)$ is an upper triangular matrix 
with $1$ on the diagonal line.  

This is a place where a connection with the KP hierarchy \cite{MJD-book} 
shows up.  Namely, the variables $\bst = (t_1,t_2,\ldots)$ 
are nothing but the ``time variables'' of the KP hierarchy, 
and the Schur functions $s_\lambda[\bst]$ are special tau functions.  
As first pointed out by Miwa \cite{Miwa82}, 
viewing the tau function as a function of the $\bsx$ variables 
leads to a discrete (or difference) analogue of the KP hierarchy. 
For this reason, the $\bsx$ variables are sometimes referred to 
as ``Miwa variables'' in the literature of integrable systems.

\subsection{Tau functions of KP hierarchy}

Let us use the notation $\tau[\bst]$ for the tau function  
in the usual sense (namely, a function of $\bst$), 
and let $\tau(\bsx)$ denote the function obtained 
from $\tau[\bst]$ by the change of variables (\ref{t-x}).  
It is the latter that plays a central role in this paper.  

A general tau function of the KP hierarchy is a linear combination 
of the Schur functions 
\beq
  \tau[\bst] 
  = \sum_{\lambda}c_\lambda s_\lambda[\bst], 
\label{KP-tau-general}
\eeq
where the coefficients $c_\lambda$ are Pl\"ucker coordinates 
of a point of an infinite dimensional Grassmann manifold 
(Sato Grassmannian) \cite{SS82}.  
Roughly speaking, the Sato Grassmannian consists of 
linear subspaces $W \simeq \CC^{\NN}$ of 
a fixed linear space $V \simeq \CC^{\ZZ}$. 
We shall not pursue those fully general tau functions 
in the following.  

We are interested in a smaller (but yet infinite dimensional) 
class of tau functions such that $c_\lambda = 0$ 
for all partitions with $l(\lambda) > N$.  
This corresponds to a submanifold $\mathrm{Gr}(N,\infty)$ 
of the full Sato Grassmannian.  The Pl\"ucker coordinates 
$c_\lambda$ are labelled by partitions of the form  
$\lambda = (\lambda_1,\ldots,\lambda_N)$, 
and given by finite determinants as 
\beq
  c_\lambda = \det(f_{i,l_j})_{i,j=1}^N, \quad
  l_i := \lambda_i - i + N. 
\eeq
Note that the sequences $\lambda_i$'s and $l_i$'s of 
non-negative integers are in one-to-one correspondence: 
\beqnn
  \infty > \lambda_1 \ge \cdots \lambda_N \ge 0 
  \quad \longleftrightarrow \quad 
  \infty > l_1 > \cdots > l_N \ge 0. 
\eeqnn
The $N \times \infty$ matrix 
\beqnn
  F = (f_{ij})_{i = 1,\ldots,N,\; j = 0,1,\ldots}
\eeqnn
of parameters represent a point of the Grassmann manifold 
$\mathrm{Gr}(N,\infty)$. 

By the Cauchy-Binet formula, the tau function  $\tau(\bsx)$ 
in the $\bsx$-picture can be expressed as 
\beq
  \tau(\bsx) 
  = \sum_{\infty>l_1>\cdots>l_N\ge 0}
    \frac{\det(f_{i,l_j})_{i,j=1}^N\det(x_i^{l_j})_{i,j=1}^N}{\Delta(\bsx)}
  = \frac{\det(f_i(x_j))_{i,j=1}^N}{\Delta(\bsx)}, 
\label{KP-tau-NxN}
\eeq
where $f_i(x)$'s are the power series of the form 
\beqnn
  f_i(x) = \sum_{l=0}^\infty f_{il}x^l.
\eeqnn

In particular, if there is a positive integer $M$ 
such that 
\beqnn
  f_{ij} = 0 \quad \mbox{for}\quad  i \ge M+N
\eeqnn
(in other words, $f_i(x)$'s are polynomials 
of degree less than $M+N$), 
the Pl\"ucker coordinate $c_\lambda$ vanishes  
for all Young diagrams not contained 
in the $N \times M$ rectangular Young diagram, 
namely, 
\beqnn
  c_\lambda = 0 \quad 
  \mbox{for}\quad \lambda \not\subseteq 
  (M^N) := (\underbrace{M,\ldots,M}_{N})
\eeqnn
The tau function $\tau[\bst]$ thereby becomes 
a linear combination of a finite number of 
Schur function, hence a polynomial in $\bst$.  
Geometrically, these solutions of the KP hierarchy 
sit on the finite dimensional Grassmann manifold 
$Gr(N,N+M)$ of the Sato Grassmannian.

\subsection{Tau functions of 2-KP hierarchy}

The tau function $\tau[\bst,\bstbar]$ 
of the 2-component KP (2-KP) hierarchy 
is a function of two sequences 
$\bst = (t_1,t_2,\ldots)$ and 
$\bstbar = (\tbar_1,\tbar_2,\ldots)$ 
of time variables, and can be expressed as 
\beq
  \tau[\bst,\bstbar]
  = \sum_{\lambda,\mu}
    c_{\lambda\mu}s_\lambda[\bst]s_\mu[\bstbar], 
\label{2KP-tau-general}
\eeq
where $c_{\lambda\mu}$'s are Pl\"ucker coordinates 
of a point of a 2-component analogue of 
the Sato Grassmannnian (which is, actually, 
isomorphic to the one-component version) \cite{SS82}. 

The aforementioned class of tau functions 
of the KP hierarchy can be generalized 
to the 2-component case.  Such tau functions 
correspond to points of the submanifold 
$\mathrm{Gr}(M+N,2\infty)$ of the 2-component 
Sato Grassmannian.   For those tau functions,  
the Pl\"ucker coordinates $c_{\lambda\mu}$ vanish 
if $l(\lambda) > M$ or $l(\mu) > N$; 
the remaining Pl\"ucker coordinates are given 
by finite determinants of a matrix with 
two rectangular blocks of size $(M+N)\times M$ 
and $(M+N)\times N$ as 
\beq
  c_{\lambda\mu} = \det(f_{i,l_j} \mid g_{i,m_k}), 
\label{2KP-c}
\eeq
where $i$ is the row index ranging over $i = 1,\ldots,M+N$ 
and $j,k$ are column indices in the two blocks 
ranging over $j = 1,\ldots,M$ and $k = 1,\ldots,N$, respectively.  
$l_j$'s and $m_k$'s are related to the parts of 
$\lambda = (\lambda_j)_{j=1}^M$ and $\mu = (\mu_i)_{j=1}^N$ as 
\beqnn
  l_j = \lambda_j - j + M, \quad 
  m_k = \mu_k - k + N.
\eeqnn

By the change of variables from $\bsx$ and $\bsy$ to 
\beq
  t_n = \frac{1}{n}\sum_{j=1}^M x_j^n, \quad 
  \tbar_n = \frac{1}{n}\sum_{k=1}^N y_k^n, 
\label{ttbar-xy}
\eeq
the tau function $\tau[\bst,\bstbar]$ is converted 
to the $(\bsx,\bsy)$-picture $\tau(\bsx,\bsy)$. 
Again by the Cauchy-Binet formula, $\tau(\bsx,\bsy)$ 
turns out to be a quotient of two determinants as 
\beq
  \tau(\bsx,\bsy) 
  = \frac{\det(f_i(x_j) \mid g_i(y_k))}{\Delta(\bsx)\Delta(\bsy)}, 
\label{2KP-tau-(M,N)block}
\eeq
where the denominator is the determinant 
with the same block structure as (\ref{2KP-c}), 
and $f_i(x)$ and $g_j(y)$ are power series of the form  
\beqnn
  f_i(x) = \sum_{l=0}^\infty f_{il}x^l,\quad 
  g_i(y) = \sum_{l=0}^\infty g_{il}y^l. 
\eeqnn

\subsection{Tau function of 2D Toda hierarchy}

The 2-KP hierarchy is closely related to the 2D Toda hierarchy 
\cite{UT84}.  The tau function $\tau_s[\bst,\bstbar]$ of 
the 2D Toda hierarchy depends on a discrete variable 
(lattice coordinate) $s$ alongside the two series 
of time variables $\bst$ and $\bstbar$.  
For each value of $s$, $\tau_s[\bst,\bstbar]$ 
is a tau function of the 2-KP hierarchy, 
and these 2-KP tau functions are mutually connected 
by a kind of B\"acklund transformations.  
Consequently, $\tau_s[\bst,\bstbar]$ can be expressed 
as shown in (\ref{2KP-tau-general}) with the coefficients 
$c_{s\lambda\mu}$ depending on $s$.  

Actually, it is more natural to use 
$s_\lambda[\bst]s_\mu[-\bstbar]$ rather than 
$s_\lambda[\bst]s_\mu[\bstbar]$ for the Schur function 
expansion of the Toda tau function \cite{Takasaki84}.  
(Note that $s_\mu[-\bstbar]$ can be rewritten as 
\beqnn
  s_\mu[-\bstbar] = (-1)^{|\mu|}s_{\tp{\mu}}[\bst], 
\eeqnn
where $\tp{\mu}$ denotes the transpose of $\mu$.) 
Expanded in these product of tau functions as
\beq
  \tau_s[\bst,\bstbar] 
  = \sum_{\lambda,\mu}c_{s\lambda\mu}s_\lambda[\bst]s_\mu[-\bstbar], 
\label{Toda-tau-general}
\eeq
the coefficients $c_{s\lambda\mu}$ become 
Pl\"ucker coordinates of an infinite dimensional 
flag manifold.  Intuitively, they are minor determinants 
\beq
  c_{s\lambda\mu} = \det(g_{\lambda_i-i+s,\mu_j-j+s})_{i,j=1}^\infty
\eeq
of an infinite matrix $g = (g_{ij})_{i,j\in\ZZ}$, 
though this definition requires justification \cite{Takasaki84}. 
In particular, if $g$ is a diagonal matrix, 
the coefficients $c_{s\lambda\mu}$ are also diagonal 
(namely, $c_{s\lambda\mu} \propto \delta_{\lambda\mu}$) 
and the Schur function expansion (\ref{Toda-tau-general}) 
simplifies to the ``diagonal'' form 
\beq
  \tau_s[\bst,\bstbar] 
  = \sum_{\lambda}c_{s\lambda}s_\lambda[\bst]s_{\lambda}[-\bstbar],\quad
  c_{s\lambda} = \prod_{i=1}^\infty g_{\lambda_i-i+s}. 
\label{Toda-tau-diagonal}
\eeq

If we reformulate the 2D Toda hierarchy 
on the semi-infinite lattice $s \ge 0$, 
the infinite determinants defining $c_\lambda$'s 
are replaced by finite determinants, 
and $\tau_s[\bst,\bstbar]$ itself becomes 
a finite determinant.  We shall encounter an example 
of such tau functions in the next section.

\section{6-vertex model with DWBC}

\subsection{Setup of model}

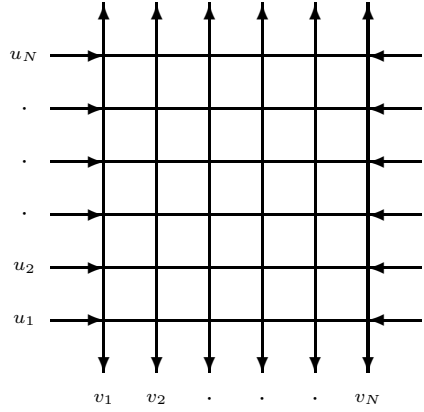
\begin{figure}
\begin{center}
\begin{picture}(150,150)(-10,-10)
\thicklines
\multiput(20,20)(0,20){6}{\line(1,0){100}}
\multiput(20,20)(20,0){6}{\line(0,1){100}}
\multiput(0,20)(0,20){6}{\vector(1,0){20}}
\multiput(140,20)(0,20){6}{\vector(-1,0){20}}
\multiput(20,20)(20,0){6}{\vector(0,-1){20}}
\multiput(20,120)(20,0){6}{\vector(0,1){20}}
\put(-10,20){\makebox(0,0){$\scriptstyle u_1$}}
\put(-10,40){\makebox(0,0){$\scriptstyle u_2$}}
\multiput(-10,60)(0,20){3}{\makebox(0,0){$\cdot$}}
\put(-10,120){\makebox(0,0){$\scriptstyle u_N$}}
\put(20,-10){\makebox(0,0){$\scriptstyle v_1$}}
\put(40,-10){\makebox(0,0){$\scriptstyle v_2$}}
\multiput(60,-10)(20,0){3}{\makebox(0,0){$\cdot$}}
\put(120,-10){\makebox(0,0){$\scriptstyle v_N$}}
\end{picture}
\end{center}
\caption{Square lattice with DWBC}
\label{fig-DWBC}
\end{figure}

We consider the 6-vertex model on an $N \times N$ 
square lattice with inhomogeneity parameters 
$\bsu = (u_1,\ldots,u_N)$ and $\bsv = (v_1,\ldots,v_N)$ 
assigned to the rows and columns.   The boundary 
of the lattice is supplemented with extra edges 
pointing outwards, and the domain-wall boundary condition 
(DWBC) is imposed on these extra edges.  Namely, 
the arrows on the extra edges on top and bottom 
of the boundary are pointing outwards, 
and those on the other extra edges are pointing inwards 
(see figure \ref{fig-DWBC}). 

The vertex at the intersection of the $i$-th row 
and the $j$-th column is given the following weight 
$w_{ij}$ determined by the configuration of arrows 
on the adjacent edges: 
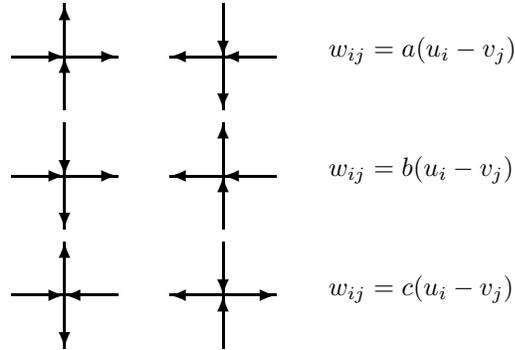
\begin{center}
\begin{picture}(140,40)
\thicklines
\put(0,20){\vector(1,0){20}}
\put(20,20){\vector(1,0){20}}
\put(20,0){\vector(0,1){20}}
\put(20,20){\vector(0,1){20}}
\put(80,20){\vector(-1,0){20}}
\put(100,20){\vector(-1,0){20}}
\put(80,20){\vector(0,-1){20}}
\put(80,40){\vector(0,-1){20}}
\put(120,20){\text{$w_{ij} = a(u_i-v_j)$}}
\end{picture}
\\
\smallskip
\begin{picture}(140,40)
\thicklines
\put(0,20){\vector(1,0){20}}
\put(20,20){\vector(1,0){20}}
\put(20,20){\vector(0,-1){20}}
\put(20,40){\vector(0,-1){20}}
\put(80,20){\vector(-1,0){20}}
\put(100,20){\vector(-1,0){20}}
\put(80,0){\vector(0,1){20}}
\put(80,20){\vector(0,1){20}}
\put(120,20){\text{$w_{ij} = b(u_i-v_j)$}}
\end{picture}
\\
\smallskip
\begin{picture}(140,40)
\thicklines
\put(0,20){\vector(1,0){20}}
\put(40,20){\vector(-1,0){20}}
\put(20,20){\vector(0,-1){20}}
\put(20,20){\vector(0,1){20}}
\put(80,20){\vector(-1,0){20}}
\put(80,20){\vector(1,0){20}}
\put(80,0){\vector(0,1){20}}
\put(80,40){\vector(0,-1){20}}
\put(120,20){\text{$w_{ij} = c(u_i-v_j)$}}
\end{picture}
\end{center}
The weight functions $a(u),b(u),c(u)$ are defined as 
\beq
  a(u) = \sinh(u+\gamma),\quad 
  b(u) = \sinh u,\quad 
  c(u) = \sinh\gamma, 
\label{abc}
\eeq
where $\gamma$ is a parameter.  
Thus the partition function of this model is defined 
as a function of the inhomogeneity parameters 
$\bsu$ and $\bsv$: 
\beqnn
  Z_N = Z_N(\bsu,\bsv) 
  = \sum_{\mbox{configuration}}\prod_{i,j=1}^N w_{ij}. 
\eeqnn

\subsection{Izergin-Korepin formula for $Z_N$}

According to the result of Korepin \cite{Korepin82} 
and Izergin \cite{Izergin87}, the partition function $Z_N$ 
has the determinant formula 
\begin{multline}
  Z_N =\frac{\prod_{i,j=1}^N\sinh(u_i-v_j+\gamma)\sinh(u_i-v_j)}
         {\prod_{1\le i<j\le N}\sinh(u_i-u_j)\sinh(v_j-v_i)}\\
  \quad\times 
    \det\left(\frac{\sinh\gamma}{\sinh(u_i-v_j+\gamma)\sinh(u_i-v_j)}
    \right)_{i,j=1}^N, 
\end{multline}
which one can rewrite as 
\begin{multline}
  Z_N =\frac{\sinh^N\eta}
         {\prod_{1\le i<j\le N}\sinh(u_i-u_j)\sinh(v_j-v_i)}\\
  \quad\times
    \det\left(\frac{\prod_{k=1}^N\sinh(u_i-v_k+\gamma)\sinh(u_i-v_k)}
    {\sinh(u_i-v_j+\gamma)\sinh(u_i-v_j)}\right)_{i,j=1}^N
\end{multline}
and 
\begin{multline}
  Z_N =\frac{\sinh^N\gamma}
         {\prod_{1\le i<j\le N}\sinh(u_i-u_j)\sinh(v_j-v_i)}\\
  \quad\times
    \det\left(\frac{\prod_{k=1}^N\sinh(u_k-v_j+\gamma)\sinh(u_k-v_j)}
    {\sinh(u_i-v_j+\gamma)\sinh(u_i-v_j)}\right)_{i,j=1}^N. 
\end{multline}
If we introduce the new variables and parameters \cite{FWZ09a} 
\beqnn
  x_i := e^{2u_i},\quad y_i := e^{2v_i},\quad q := e^{-\gamma}, 
\eeqnn
we can rewrite these formulae as 
\begin{multline}
  Z_N = C_N\prod_{i,j=1}^N(x_iq^{-1}-y_jq)(x_i-y_j)\\
  \quad\times
    \frac{1}{\Delta(\bsx)\Delta(\bsy)}
    \det\left(\frac{q^{-1}-q}{(x_iq^{-1}-y_jq)(x_i-y_j)}\right)_{i,j=1}^N,
\label{rat-IK0}
\end{multline}
\beq
  Z_N = \frac{C_N(q^{-1}-q)^N}{\Delta(\bsx)\Delta(\bsy)}
    \det\left(\frac{\prod_{k=1}^N(x_iq^{-1}-y_kq)(x_i-y_k)}
       {(x_iq^{-1}-y_jq)(x_i-y_j)}\right)_{i,j=1}^N, 
\label{rat-IK1}
\eeq
and 
\beq
  Z_N = \frac{C_N(q^{-1}-q)^N}{\Delta(\bsx)\Delta(\bsy)}
    \det\left(\frac{\prod_{k=1}^N(x_kq^{-1}-y_jq)(x_k-y_j)}
       {(x_iq^{-1}-y_jq)(x_i-y_j)}\right)_{i,j=1}^N, 
\label{rat-IK2}
\eeq
where $C_N = C_N(\bsu,\bsv)$ is an exponential function 
of a linear combination of $u_i$'s and $v_i$'s.  
Apart from this simple factor, $Z_N$ thus reduces 
to a rational function of $\bsx = (x_1,\ldots,x_N)$ 
and $\bsy = (y_1,\ldots,y_N)$.

\subsection{KP and 2-KP tau functions hidden in $Z_N$}

As pointed out by Foda et al. \cite{FWZ09a}, 
two KP tau functions are hidden 
in these determinant formulae of $Z_N$. 
Firstly, if $y_i$'s are considered to be constants, 
the $\bsx$-dependent part of (\ref{rat-IK1}) 
give the function 
\beq
  \tau_1(\bsx) = \frac{\det(f_j(x_i))_{i,j=1}^N}{\Delta(\bsx)}, \quad
  f_j(x) := \frac{\prod_{k=1}^N(xq^{-1}-y_kq)(x-y_k)}{(xq^{-1}-y_jq)(x-y_j)}. 
\eeq
This is a tau function of the KP hierarchy 
with respect to $t_n = \frac{1}{n}\sum_{k=1}^Nx_k^n$.  
Moreover, since $f_j(x)$'s are polynomials in $x$, 
this tau function is as a polynomial in $\bst$.  
In the same sense, if $x_i$'s are considered 
to be constants, the $\bsy$-dependent part of 
(\ref{rat-IK2}) gives the function 
\beq
  \tau_2(\bsy) = \frac{\det(g_i(y_j))_{i,j=1}^N}{\Delta(\bsy)},\quad 
  g_i(y) := \frac{\prod_{k=1}^N(x_iq^{-1}-yq)(x_i-y)}{(x_iq^{-1}-yq)(x_i-y)}, 
\eeq
which is a polynomial tau function of the KP hierarchy 
with respect to $\tbar_n = \frac{1}{n}\sum_{k=1}^Ny_k^n$.  

Thus, apart from an irrelevant factor,  
$Z_N$ is a tau function of the KP hierarchy 
with respect to $\bsx$ and $\bsy$ separately.  
It will be natural to ask whether a tau function 
of the 2-KP hierarchy is hidden in $Z_N$. 
 
A partial answer can be found in the work of 
Stroganov \cite{Stroganov06} and Okada \cite{Okada06}.   
According to their results, if $q = e^{\pi i/3}$, 
the partition function coincides, 
up to a simple factor, with a single Schur function 
of $(\bsx,\bsy)$ as 
\beq
  Z_N = (\mbox{simple factor})s_\lambda(\bsx,\bsy), 
\label{SO-formula}
\eeq
where $\lambda$ is the double staircase partition 
\beqnn
  \lambda = (N-1,N-1,N-2,N-2,\ldots,1,1) 
\eeqnn
of length $2N$.  By the way, for any partition $\lambda 
= (\lambda_1,\ldots,\lambda_{2N})$ of length $\le 2N$, 
the Weyl character formula for $s_\lambda(\bsx,\bsy)$ reads 
\beqnn
  s_\lambda(\bsx,\bsy) 
  = \frac{\det(x_j^{\lambda_i-i+2N} \mid y_k^{\lambda_i-i+2N})}
    {\Delta(\bsx,\bsy)}, 
\eeqnn
where the row index $i$ ranges over $i = 1,\ldots,2N$ 
and the column indices $j,k$ in the two blocks 
over $j,k = 1,\ldots,N$.  Multiplying this function 
by $\Delta(\bsx,\bsy)/\Delta(\bsx)\Delta(\bsy)$ gives 
\beq
  \frac{\Delta(\bsx,\bsy)}{\Delta(\bsx)\Delta(\bsy)}s_\lambda(\bsx,\bsy) 
  = \frac{\det(x_j^{\lambda_i-i+2N} \mid y_k^{\lambda_i-i+2N})}
    {\Delta(\bsx)\Delta(\bsy)}, 
\eeq
which may be thought of as a 2-KP tau function 
of the form (\ref{2KP-tau-(M,N)block}).  

Another answer, which is valid for arbitrary values of $q$, 
was found by Zinn-Justin (private communication).  
Let us note that (\ref{rat-IK0}) can be rewritten as 
\begin{multline}
  Z_N 
= C_N(q^{-1}-q)^N\prod_{i,j=1}^N(1-q^{-2}x_iy_j^{-1})(1-x_iy_j^{-1})
  \prod_{1\le i<j\le N}(- y_iy_j)\\
  \quad\times
    \frac{1}{\Delta(\bsx)\Delta(\bsy^{-1})}
    \det\left(\frac{1}{(1-q^{-2}x_iy_j^{-1})(1-x_iy_j^{-1})}\right)_{i,j=1}^N,
\label{ZJ}
\end{multline}
where 
\beqnn
  \bsy^{-1} = (y_1^{-1},\ldots,y_N^{-1}). 
\eeqnn
The last part of thins expression, namely 
the quotient of the determinant by 
the Vandermonde determinants $\Delta(\bsx)\Delta(\bsy^{-1})$, 
may be thought of as a 2-KP tau function with respect to 
the time variables 
\beq
  t_n = \frac{1}{n}\sum_{k=1}^N x_k^n, \quad 
  \tbar_n = - \frac{1}{n}\sum_{k=1}^N y_k^{-n}. 
\label{Toda-ttbar-xy}
\eeq
This is a special case of the tau functions 
\beq 
  \tau(\bsx,\bsy) 
  = \frac{\det(h(x_iy_j^{-1}))_{i,j=1}^N}{\Delta(\bsx)\Delta(\bsy^{-1})}, 
\label{OS}
\eeq
considered by Orlov and Shiota \cite{OS05}, 
where $h(z)$ is an arbitrary power series of the form 
\beqnn
  h(z) = \sum_{n=0}^\infty h_nz^n,\quad 
  h_n \not= 0 \quad \mbox{for} \quad n \ge 0. 
\eeqnn
By the Cauchy-Binet formula, 
$\tau(\bsx,\bsy)$ can be expanded as 
\beq
  \tau(\bsx,\bsy) = \sum_{\lambda = (\lambda_1,\ldots,\lambda_N)}
    c_\lambda s_\lambda(\bsx)s_\lambda(\bsy^{-1}), \quad 
  c_\lambda := \prod_{i=1}^N h_{\lambda_i-i+N}. 
\label{OS-expansion}
\eeq
This is an analogue (for a semi-infinite lattice) 
of the Toda tau functions of the diagonal form 
(\ref{Toda-tau-diagonal}).  Note that the role 
of the lattice coordinate $s$ is played by $N$. 
Since the number of the Miwa variables in 
(\ref{Toda-ttbar-xy}) also depends on $N$, 
translation to the lanugage of the 2D Toda hierarchy 
is somewhat tricky, but this tricky situation 
is rather common in random matrix models \cite{OS05}. 
Thus, though not of the type shown in (\ref{2KP-tau-(M,N)block}), 
the last part of (\ref{ZJ}) turns out to be a 2-KP tau function.  

Lastly, let us mention that Korepin and Zinn-Justin 
considered the partition function in the homogeneous limit 
as $u_i,v_j \to 0$ \cite{KZJ00}.  In that limit, 
the partition function reduces, up to a simple factor, 
to a special tau function of the 1D Toda equation, 
and can be treated as an analogue of random matrix models.

\section{Scalar product of states in finite XXZ spin chain}

\subsection{$L$- and $T$-matrices for spin $1/2$ chain}

We consider a finite XXZ spin chain of spin $1/2$ and length $N$ 
with inhomongeneity parameters $\xi_l$, $l = 1,\ldots,N$.  
To define local $L$-matrices, let us introduce 
the $2 \times 2$ matrix $L(u) = (L_{ij}(u))_{i,j=1,2}$ 
of the $2 \times 2$ blocks 
\beqnn
  L_{11}(u) = a(u)\frac{1+\sigma^3}{2} + b(u)\frac{1-\sigma^3}{2},\quad
  L_{12}(u) = c(u)\sigma^{-},\\
  L_{21}(u) = c(u)\sigma^{+},\quad
  L_{22}(u) = b(u)\frac{1+\sigma^3}{2} + a(u)\frac{1-\sigma^3}{2},  
\eeqnn
where $\sigma^{\pm}$ and $\sigma^3$ are the Pauli matrices 
\beqnn
  \sigma^{+} 
  = \left(\begin{array}{cc}
    0 & 1\\
    0 & 0
    \end{array}\right),\quad 
  \sigma^{-}
  = \left(\begin{array}{cc}
    0 & 0\\
    1 & 0
    \end{array}\right),\quad 
  \sigma^3 
  = \left(\begin{array}{cc}
    1 & 0\\
    0 & -1
    \end{array}\right).
\eeqnn
These $2 \times 2$ blocks are understood to act  
on the single spin space $\CC^2$.   
The structure functions $a(u),b(u),c(u)$ are the same as 
the weight functions (\ref{abc}) for the 6-vertex model, 
and built into the $R$-matrix 
\beqnn
  R(u-v) 
  = \left(\begin{array}{cccc}
    a(u-v) & 0 & 0 & 0 \\
    0 & b(u-v) & c(u-v) & 0 \\
    0 & c(u-v) & b(u-v) & 0 \\
    0 & 0 & 0 & a(u-v)
    \end{array}\right). 
\eeqnn

Let $L^{(l)}(u-\xi_l) = (L^{(l)}_{ij}(u-\xi))_{i,j=1,2}$ 
be the local $L$-matrix at the $l$-th site, 
\beqnn
  L^{(l)}_{ij}(u - \xi_l) 
  = \cdots \otimes 1 \otimes L_{ij}(u - \xi_l) 
    \otimes 1 \otimes \cdots, 
\eeqnn
and define the $T$-matrix as 
\beqnn
  T(u)
  = \left(\begin{array}{cc}
    A(u) & B(u) \\
    C(u) & D(u)
    \end{array}\right)
  = L^{(1)}(u - \xi_1)\cdots L^{(N)}(u - \xi_N). 
\eeqnn
The matrix elements of these matrices, hence 
the trace of the $T$-matrix 
\beqnn
  \calT(u) = \Tr T(u) = A(u) + D(u) 
\eeqnn
as well, are operators on the full spin space 
$V = \bigotimes_{l=1}^N\CC^2$.  
The $L$-matrices satisfy the local intertwining relations 
\begin{multline}
  R(u-v)(L^{(l)}(u)\otimes I)(I\otimes L^{(m)}(v)) \\
= (I\otimes L^{(m)}(v))(L^{(l)}(u)\otimes I)R(u-v), 
\end{multline}
where $R(u-v)$, $L^{(l)}(u) \otimes I$ and 
$I \otimes K^{(m)}(v)$ are understood to be 
$4 \times 4$ matrices (of scalars or of spin operators 
on $V$) acting on the tensor product $\CC^2 \otimes \CC^2$ 
of two copies of the auxiliary space $\CC^2$.  
These local intertwining relations lead to 
the global intertwining relation 
\beq
  R(u-v)(T(u)\otimes I)(I\otimes T(v)) 
= (I\otimes T(v))(T(u)\otimes I)R(u-v) 
\label{RTT=TTR}
\eeq
for the $T$-matrix.  This is a compact expression 
of many bilinear relations among the matrix elements 
of $T(u)$ and $T(v)$, such as 
\beq
\begin{aligned}
  A(u)B(v) &= f(u-v)B(v)A(u) - g(u-v)B(u)A(v),\\
  D(u)B(v) &= f(u-v)B(v)D(u) - g(u-v)B(u)D(v),\\
  C(u)B(v) &= g(u-v)(A(u)D(v)  - A(v)D(u)) 
\end{aligned}
\eeq
and 
\beq
\begin{aligned}
{} [A(u),A(v)] = 0, &\quad [B(u),B(v)] = 0,\\
   [C(u),C(v)] = 0, &\quad [D(u),D(v)] = 0, 
\end{aligned}
\eeq
where 
\beqnn
  f(u) = \frac{a(u)}{b(u)} = \frac{\sinh(u+\gamma)}{\sinh u}, \quad
  g(u) = \frac{c(u)}{b(u)} = \frac{\sinh \gamma}{\sinh u}. 
\eeqnn
A consequence of those relations 
is the fact that $\calT(u)$ and $\calT(v)$ commute 
for any values of $u,v$: 
\beq
  [\calT(u), \calT(v)] = 0. 
\eeq
The algebraic Bethe ansatz is a method for constructing 
simultaneous eigenstates (called ``Bethe states'') 
of $\calT(u)$ for all values of $u$.

\subsection{Algebraic Bethe ansatz}

Let us introduce the pseudo-vacuum $0\rangle$ 
and its dual $\langle 0|$: 
\beqnn
  \langle 0|
  = \bigotimes_{l=1}^N
    \left(\begin{array}{cc} 1 & 0 \end{array}\right) \in V^*, 
  \quad 
  |0\rangle 
  = \bigotimes_{l=1}^N 
    \left(\begin{array}{c} 1 \\ 0 \end{array}\right) \in V.
\eeqnn
They indeed satisfy the vacuum conditions 
\beq
\begin{aligned}
  \langle 0|B(u) &= 0,& 
  C(u)|0\rangle &= 0,\\
  A(u)|0\rangle &= \alpha(u)|0\rangle,&
  D(u)|0\rangle &= \delta(u)|0\rangle,\\
  \langle 0|A(u) &= \alpha(u)\langle 0|,&
  \langle 0|D(u) &= \delta(u)\langle 0|,
\end{aligned}
\eeq
where 
\beqnn
  \alpha(u) = \prod_{l=1}^N \sinh(u - \xi_l + \gamma),\quad
  \delta(u) = \prod_{l=1}^N \sinh(u - \xi_l). 
\eeqnn
For notational convenience, we introduce 
the reflection coefficients 
\beqnn
  r(u) = \frac{\alpha(u)}{\delta(u)}. 
\eeqnn

Bethe states are generated from $|0\rangle$ 
by the action of $B(v_j)$'s.  Suppose that $v_j$'s 
satisfy the Bethe equations
\beq
  r(v_i)\prod_{j\not= i}
        \frac{\sinh(v_i-v_j-\gamma)}{\sinh(v_i-v_j+\gamma)}
  = 1, \quad i = 1,\ldots,n. 
\label{Bethe-eq}
\eeq
The state $\prod_{i=1}^nB(v_i)|0\rangle$ then becomes 
an eigenstate of $\calT(u)$: 
\begin{multline}
  \calT(u)\prod_{i=1}^nB(v_i)|0\rangle \\
  = \left(\alpha(u)\prod_{i=1}^nf(v_i-u) 
      + \delta(u)\prod_{i=1}^nf(u-v_i)\right)
    \prod_{i=1}^nB(v_i)|0\rangle. 
\end{multline}

Let us remark that the operators $A(u),B(u),C(u),D(u)$ 
are related to a row-to-row transfer matrix 
of the 6-vertex model on the square lattice.  
One can thereby derive the identities \cite{ICK91} 
\beqnn
  \langle 0|\prod_{i=1}^N C(u_i) 
  = \langle \zerobar|Z_N(u_1,\ldots,u_N,\xi_1,\ldots,\xi_N),\\
  \prod_{i=1}^NB(u_i)|0\rangle 
  = Z_N(u_1,\ldots,u_N,\xi_1,\ldots,\xi_N)|\zerobar\rangle,
\eeqnn
where $u_i$'s are free (namely, not required to satisfy 
the Bethe equations) variables, and $|\zerobar\rangle$ 
and  $\langle\zerobar|$ denote the ``anti-pseudo-vacuum'' 
and its dual: 
\beqnn
  \langle\zerobar| 
  = \bigotimes_{l=1}^N
    \left(\begin{array}{cc} 0 & 1\end{array}\right) \in V^*,\quad
  |\zerobar\rangle 
  = \bigotimes_{l=1}^N 
    \left(\begin{array}{c} 0 \\  1\end{array}\right) \in V.
\eeqnn

\subsection{Slavnov formula for scalar product}

Let $\bsu = (u_1,\ldots,u_n)$ be free variables 
and $\bsv = (v_1,\ldots,v_n)$ satisfy 
the Bethe equations (\ref{Bethe-eq}).   
The Slavnov formula \cite{Slavnov89,KMT99} 
for the scalar product 
\beqnn
  S_n(\bsu,\bsv) 
  = \langle 0|\prod_{i=1}^nC(u_i)\prod_{i=1}^nB(v_i)|0\rangle
\eeqnn
reads 
\beq
  S_n(\bsu,\bsv) 
  = \frac{\prod_{i=1}^n\delta(u_i)\delta(v_i) 
          \prod_{i,j=1}^n\sinh(u_i-v_j+\gamma)}
    {\prod_{1\le i<j\le n}\sinh(u_i-u_j)\sinh(v_j-v_i)} 
    \det(H_{ij})_{i,j=1}^n, 
\eeq
where 
\beqnn
  H_{ij} 
  = \frac{\sinh\gamma}{\sinh(u_i-v_j+\gamma)\sinh(u_i-v_j)}
    \left(1 
      - r(u_i)\prod_{k\not= i}
        \frac{\sinh(u_i-v_k-\gamma)}{\sinh(u_i-v_k+\gamma)}\right).
\eeqnn
One can rewrite this formula as 
\beq
  S_n(\bsu,\bsv) 
  = \frac{\sinh^n\gamma\prod_{i=1}^n\delta(v_i)}
    {\prod_{1\le i<j\le n}\sinh(u_i-u_j)\sinh(v_j-v_i)} 
    \det(K_{ij})_{i,j=1}^n, 
\eeq
where 
\beqnn
  K_{ij} 
  = \frac{\displaystyle 
          \delta(u_i)\prod_{k\not=j}\sinh(u_i-v_k+\gamma)
        - \alpha(u_i)\prod_{k\not=j}\sinh(u_i-v_k-\gamma)}
         {\sinh(u_i-v_j)}. 
\eeqnn

\subsection{KP tau function hidden in $S_n(\bsu,\bsv)$}

If we introduce the new variables and parameters 
\cite{FWZ09b}
\beqnn
  x_i := e^{2u_i},\quad y_i := e^{2v_i},\quad 
  z_i := e^{2\xi_i},\quad q := e^{-\gamma}, 
\eeqnn
the Slavnov formula can be converted 
to the almost rational form 
\beq
  S_n(\bsu,\bsv) 
  = \frac{C_n\sinh^n\gamma\prod_{i=1}^n\delta(v_i)}{\Delta(\bsy)}
    \frac{\det(f_j(x_i))_{i,j=1}^n}{\Delta(\bsx)},
\eeq
where $C_n = C_n(\bsu,\bsv)$ is an exponential function 
of a linear combination of $u_i$'s and $v_i$'s, and 
\beqnn
  f_j(x) 
  = \frac{\displaystyle 
       \prod_{l=1}^N(x - z_l)\prod_{k\not= j}(q^{-1}x - qy_k)
      - \prod_{l=1}^N(q^{-1}x - qz_l)\prod_{k\not= j}(qx - q^{-1}y_k)}
    {x - y_j}. 
\eeqnn
Thus, as pointed out by Foda et al. \cite{FWZ09b}, 
a KP tau function of the form (\ref{KP-tau-NxN}) 
is hidden in $S(\bsu,\bsv)$.  Moreover, 
since the Bethe equations (\ref{Bethe-eq}) 
imply the equations 
\begin{multline*}
    \prod_{l=1}^N(y_i - z_l)\prod_{k\not= j}(q^{-1}y_i - qy_k)\\
  = \prod_{l=1}^N(q^{-1}y_i - qz_l)\prod_{k\not= j}(qy_i - q^{-1}y_k)
  \quad (i = 1,\ldots,n) 
\end{multline*}
for $y_i$'s, the numerator of $f_j(x)$ can be 
factored out by the denominator $x - y_j$.  
Thus $f_j(x)$'s turn out to be polynomials in $x$.

\section{Scalar product of states in models at $q = 0$}

A class of solvable models, such as the phase model \cite{BIK97} 
and and the totally asymmetric simple exclusion process 
(TASEP) model \cite{Bogoliubov09}, can be formulated 
by a set of $2 \times 2$ $L$-matrices $L^{(l)}(u)$, 
$l = 1,2,\ldots,N$, and an $R$-matrix of the form 
\beqnn
R(u-v) 
  = \left(\begin{array}{cccc}
    f(u-v) & 0 & 0 & 0 \\
    0 & 1 & g(u-v) & 0 \\
    0 & g(u-v) & 0 & 0 \\
    0 & 0 & 0 & f(u-v)
    \end{array}\right), 
\eeqnn
where 
\beqnn
  f(u-v) = \frac{u^2}{u^2 - v^2}, \quad 
  g(u-v) = \frac{uv}{u^2 - v^2}.
\eeqnn
This $R$-matrix is obtained as a ``crystal'' (namely, 
$q \to 0$) limit of the $R$-matrix of the 6-vertex model 
and the XXZ spin chain.  Unlike the XXZ chain of spin $1/2$, 
the $L$-matrices are not given by $2 \times 2$ blocks 
of the $R$-matrix and take a model-dependent form.  
We define the $T$-matrix as 
\beqnn
  T(u)   T(u)
  = \left(\begin{array}{cc}
    A(u) & B(u) \\
    C(u) & D(u)
    \end{array}\right)
  = L^{(1)}(u)\cdots L^{(N)}(u) 
\eeqnn
and consider the scalar product 
\beqnn
  S_n(\bsu,\bsv) 
  = \langle 0|\prod_{i=1}^nC(u_i)\prod_{i=1}^nB(v_i)|0\rangle. 
\eeqnn

Remarkably, even if both $\bsu = (u_1,\ldots,u_n)$ 
and $\bsv = (v_1,\ldots,v_n)$ are free variables, 
the scalar product for those models 
has a determinant formula \cite{KBI-book} of the form 
\beq
  S_n(\bsu,\bsv) 
  = (\mbox{simple factor})
    \prod_{1\le i<j\le n}\frac{u_iu_j}{u_i^2-u_j^2}
    \frac{v_jv_i}{v_j^2-v_i^2}
    \det(K_{ij})_{i,j=1}^n,
\eeq
where 
\beqnn
  K_{ij} 
  = \frac{\alpha(u_i)\delta(v_j)(v_j/u_i)^{n-1} 
        - \delta(u_i)\alpha(v_j)(u_i/v_j)^{n-1}}{(u_i^2 - v_j^2)/u_iv_j},  
\eeqnn
$\alpha(u)$ and $\delta(u)$ being determined by 
the action of $A(u)$ and $D(u)$ on the pseudo-vacuum.  
If $\alpha(u)$ and $\delta(u)$ are given explicitly, 
we will be able to obtain a KP tau function 
(and hopefully a 2-KP tau function as well). 

Such an interpretation can be found most clearly 
in the cases of the phase model 
\cite{Bogoliubov05,SU05,Tsilevich05} 
and the 4-vertex model \cite{Bogoliubov07}, 
both of which are related to enumeration of 
boxed plane partitions.  The scalar product 
of Bethe states in these cases becomes, 
up to a simple factor, a sum of products 
of two Schur functions: 
\beq
  S(\bsu,\bsv) 
  = (\mbox{simple factor})
    \sum_{\lambda\subseteq (N^n)}
    s_\lambda(u_1^2,\ldots,u_n^2)
    s_\lambda(v_1^{-2},\ldots,v_n^{-2}). 
\label{PPtau-expansion}
\eeq
By the same reasoning as the interpretation of 
(\ref{OS-expansion}), one can see that 
this sum is a tau function of the 2-KP hierarchy 
(or, rather, the 2D Toda hierarchy as Zuparic argued 
\cite{Zuparic09}) with respect to the time variables 
\beqnn
  t_n = \frac{1}{n}\sum_{i=1}^n u_i^{2n},\quad
  \tbar_n = - \frac{1}{n}\sum_{i=1}^n v_i^{-2n}.
\eeqnn

Actually, this case admits yet another interpretation.  
In the course of deriving (\ref{PPtau-expansion}), 
the scalar product is shown to be a generating function 
for counting plane partitions in a box of size 
$n \times n \times N$.  It is more or less well known 
\cite{Zinn-Justin09} that the Schur function $s_{(N^n)}$ 
gives such a generating function.  
Thus, in terms of this Schur function, 
the scalar product can be expressed as 
\beq
  S(\bsu,\bsv) 
  = (\mbox{simple factor})
    s_{(N^n)}(u_1^2,\ldots,u_N^2,v_1^2,\ldots,v_N^2). 
\eeq
This is similar to the formula (\ref{SO-formula}) 
of the partition function of the 6-vertex model 
for $q = e^{\pi/3}$; one can thereby find 
an interpretation as a KP or 2-KP tau function 
with respect to the time variables 
\beqnn
  t_n = \frac{1}{n}\sum_{i=1}^n u_i^{2n},\quad
  \tbar_n = \frac{1}{n}\sum_{i=1}^n v_i^{2n}.
\eeqnn

\section*{Acknowledgments}
The author thanks Omar Foda and Paul Zinn-Justin 
for useful comments and discussion.  
This work is partly supported by Grant-in-Aid for 
Scientific Research No. 19540179 and No. 21540218 
from the Japan Society for the Promotion of Science.

\end{document}